\documentstyle[12pt,epsf,amstex]{article}
\begin{document}
\setlength{\unitlength}{1mm}
\textwidth 15.0 true cm 
\headheight 0 cm
\headsep 0 cm 
\topmargin 0.4 true in
\oddsidemargin 0.25 true in
\input epsf

\newcommand{\beq}{\begin{equation}}
\newcommand{\eeq}{\end{equation}}
\newcommand{\be}{\begin{eqnarray}}
\newcommand{\ee}{\end{eqnarray}}
\renewcommand{\vec}[1]{{\bf #1}}
\newcommand{\vecg}[1]{\mbox{\boldmath $#1$}}
\newcommand{\grpicture}[1]
{
    \begin{center}
        \epsfxsize=200pt
        \epsfysize=0pt
        \vspace{-5mm}
        \parbox{\epsfxsize}{\epsffile{#1.eps}}
        \vspace{5mm}
    \end{center}
}

\begin{flushright}

SUBATECH--2002--10

\end{flushright}

\vspace{0.5cm}

\begin{center}

{\Large\bf  Quantum Gravity as Escher's Dragon }
\footnote{Invited talk at the conference ``Modern trends
in classical approaches'' devoted to the 80-th birthday
of K.A. Ter--Martirosyan, Moscow, September 2002.} 

\vspace{1cm}

{\Large A.V. Smilga} \\

\vspace{0.5cm}

{\it SUBATECH, Universit\'e de
Nantes,  4 rue Alfred Kastler, BP 20722, Nantes  44307, France. }
\footnote{On leave of absence from ITEP, Moscow, Russia.}\\
\end{center}

\bigskip

\begin{abstract}
 The main obstacle in attempts to construct a consistent quantum
gravity is the absence of independent flat time. This can  
in principle be cured by going out to higher dimensions. 
The modern paradigm assumes that the fundamental theory of
everything is some form of string theory living in space
of more than four dimensions. We advocate another possibility
that the fundamental theory is a form of $D=4$ higher--derivative 
gravity. This class of theories
 has a nice feature of renormalizability so that
perturbative calculations are feasible. There are also
{\it finite}  ${\cal N} =4$ supersymmetric conformal supergravity
theories. This possibility is particularly attractive.
Einstein's gravity is 
obtained in a natural way as an effective low--energy theory.

The ${\cal N} =1$ supersymmetric version of the theory  has a 
natural higher--dimensional interpretation due to Ogievetsky
and Sokatchev, which involves embedding of
our curved Minkowsky space-time
manifold  into  flat 8-dimensional space. Assuming that
a variant of the finite  ${\cal N} =4$ theory also admit a similar
interpretation, this may eventually
allow one to construct  consistent quantum theory of gravity.

We argue, however, that even though future gravity theory 
will probably use higher dimensions as construction scaffolds,
its physical content and meaning should refer to 4 dimensions
where observer lives. 
\end{abstract}

\section{Introduction.}

Karen Avetovitch belongs to the first generation of the Landau
school. A characteristic feature of Landau and his disciples
was dislike of  ``philosophy''. The latter  was
understood  in broad sense as any kind of  discussion 
without explicit formulas or numbers.
A scientific paper should involve a derivation of some new formula
or new number --- this was the main lesson which Landau taught
to K.A. and which K.A. taught to his students including myself.
 In my own scientific activity, I mostly tried to follow this
commandment, but human beings are weak and sinful, and 
cannot really be Good all the time. Sometimes, when the task
to derive things scientifically is too hard (as it is the case
for quantum gravity), it is very difficult to resist the 
 temptation to think and, which is
worse,  to {\it talk} about these matters.
When discussing the foundations of quantum gravity, one {\it has}
to do a philosophical talk or no talk at all. Today I've chosen
the first option and can only hope that 
  K.A. will not condemn me too much. 

Actually, we do not understand what quantum gravity is. 
To understand why we do not understand this, let me briefly remind
the things that we understand well.
\begin{itemize}
\item We understand well Newton's laws and,
 generically, the dynamics
of any classical system where equations of motion have Cauchy
form: you set up the initial conditions at a given time moment and
find out how the system will look like at later times. The number
of dynamic variables can be finite 
(this is called classical mechanics) or continuously infinite
(this is called classical field theory).  
Such dynamic systems often enjoy 
extra symmetries. The symmetries might be global (with N\"other
currents, etc) or dynamical (involving the Hamiltonian). 
The important representative of the latter is Lorentz symmetry.
 There are also gauge symmetries, which {\it are} not symmetries but
rather additional constraints imposed in  phase space, which are
  respected during the time evolution of the system prescribed by
its Hamiltonian.
\item We know how to construct quantum counterparts for all
theories mentioned above. You introduce Hilbert space and write
the Schr\"odinger equation for wave functions 
(in case when the number
of degrees of freedom is finite) or wave functionals (in case
when the number of degrees of freedom is continuous).  
To tackle with the continuous number of dynamic variables in field
theories, one should first make it finite (introduce
ultraviolet and  infrared regularization) and then explore the limit when the corresponding cutoffs are lifted. In some cases 
(like for QED or for $\lambda \phi^4$ theory or for { any} field
theory with space-time dimension 5 or greater), 
this leads to a trouble: the continuum limit does not exist. But in
many physically important cases ($D=4$ non--Abelian gauge theories),
the continuous limit is well defined.
  \end{itemize}

And this is all that we know {\it for sure}. The reader might be
surprised why  did I not mention { classical} gravity. A 
common believe is that though quantum gravity is, indeed, 
not constructed and not understood yet, the {\it classical} 
 theory,  Einstein's gravity, is something which we know well and
are sure about.
Mostly, this is true, but not quite. The discussion of
this nontrivial point is what I would like
to begin with.

\section{Einstein's gravity.}

The action of the theory is
 \be
\label{SEinst}
S \ =\ m_P^2 \int R \sqrt{-g} \ d^4 x \ +\ 
\int {\cal L}_{\rm matter} \sqrt{-g} \ d^4 x \ .
  \ee
The equations of motion are 
   \be
\label{urEinst}
R_{\mu\nu} - \frac 12 g_{\mu\nu}R = \frac 1{m_P^2} T_{\mu\nu}\ ,
  \ee
where $R$ is scalar curvature, $R_{\mu\nu}$ --- Ricci's tensor,
$m_P$ --- Planck mass, and $T_{\mu\nu}$ - energy-momentum tensor
of the matter fields.

 The main problem of this theory is the {\it problem   of time}
(see e.g. Ref.\cite{Isham} for an extensive discussion). In 
``normal'' systems, time is an independent variable, not a dynamical
one. In gravity, time is just one of the coordinates on a $D=4$
manifold and is intertwined with spatial coordinates, which {are}
related to the  dynamic variables. The dependence on time cannot
be disentangled from other dependencies. At the classical level,
this means that the problem of soving 
the Einstein's equations (\ref{urEinst}) 
cannot be always reduced to a Cauchy problem.

We hasten to comment that, in all cases representing  physical
interest, it can. This can be done if the 4--dimensional manifold
can be represented as a set of three--dimensional slices of the 
same topology (the interval between any two points on such a slice
is space-like). In other words, the topology of space--time 
should be
$\Sigma \times R$. In the physically interesting case, $\Sigma$  
is topologically equivalent to $R^3$ and is asymptotically flat.
 Choosing {\it some} coordinate along the timelike factor $R$,
we may call it time and rewrite  Einstein's equations such that
they would express evolution with respect to this time. This 
procedure is called canonical Arnowitt-Deser-Misner formalism
\cite{ADM}.

The trouble strikes back in the following way. Suppose we 
pose some initial conditions at the spacelike slice $\Sigma$
corresponding to the moment $t=0$ 
and are interested what happens at later times. For generic initial
conditions, singularities will develop (black holes will be formed).
The formation of black holes as such does not lead to 
inconsistencies. The matter is that the singularity in the center
of the hole is normaly surrounded by an event
  horizon (as is the case
for the Schwarzschild solution) and is thereby unreachable: if we
place the observer far away from the holes, where the metric
is nearly flat, he will not get signals from the regions close
to singularities and, as far as this observer is concerned, the
future evolution of the system {\it is} uniquely determined by the
Cauchy data in the past.

The conjecture of R. Penrose was that singularities are always 
surrounded 
by  horizons and a ``naked'' singularity
is never possible (the so-called {\sl Cosmic 
censorship principle})
 \cite{Penrose}. It was found, however, that this conjecture is not
true in its strong form: there {\it are} solutions to  Einstein's 
equations involving naked singularities (see  Ref.\cite{naked}
for recent review). 
A separate  question is whether
these solutions are physically realized. The answer to this 
is probably
negative: all such solutions seem to be unstable so that a small
fluctuation of initial conditions destroy them. But {\it in
principle}, naked singularities are not forbidden
 in general relativity.

The presence of a naked singularity means that a distant observer
receives information from regions of arbitrary large curvature where
classical theory does not apply. Still, he does not receive in this
case information from the singularity proper, and Cauchy 
interpretation is not spoiled yet on this stage.
 But there are cases when it {\it is}. 
First of all, the symmetry of the equations with respect
to time reversal tells one that, 
on top of black hole solutions, there are white hole
solutions, for which the world lines, matter, and information 
flow out of the 
singularity through the horizon to infinity. Again, these solutions
are not stable and are not physically realized (at least, at the
macroscale), but, at the
foundational level, they present a trouble.

Even more this
 refers to the wormhole solutions with closed time loops
\cite{CTL}. They have roughly the same status
as the naked singularity solutions and 
white hole solutions. The topology of the corresponding 4--manifolds
is more complicated than $\Sigma \times R$ (so that
the ADM canonical formalism does not apply here) and 
involves a ``handle'' with two ``mouths''. The distance between
the mouths in the usual space may be large while the geodesic
distance measured through the wormhole may be small. 
As a result,
the particles travelling through the wormhole will effectively
move faster
than light from the viewpoint of an  outer space observer, 
and this means violation of 
causality, which {\it is} a  trouble. In particular, no Cauchy
 interpretation for the equations of motion 
is possible in this case \cite{Novikov}.

In other words, general relativity describes well observable
physical events at macroscale, but it has inherent 
problems at the foundational level. The same difficulty appears in
any gravity theory including general covariance principle. The basic
reason for this is the absence of independent flat time.

\section{Quantization}
If the problems are there at the classical level, they are not
going to disappear when we try to quantize the theory. Actually,
they become much more severe. If in the classical case
 non--causality
showed up only for rather special solutions, it is an inherent
and unavoidable feauture of quantum gravity.

I mean here in the first place Hawking's paradox \cite{Hawking}
associated
with black hole formation. As was discussed above, in 
the classical theory, there are ``benign'' solutions,
which  describe the formation of black holes prudently 
surrounded by a horizon. These solutions present no conceptual
problems. But in  quantum theory, black holes are not 
completely black, they radiate by Hawking mechanism. This 
radiation is purely stochastic and does { not} carry any
information on what particular kind of 
matter fell in the black hole.
This information is lost completely. Therefore, our system,
having 
presented a pure quantum state at $t=0$, is {\it necessarily}
transformed into mixed state after the black hole was formed
and radiated a little bit. This means loss
of unitarity\footnote{ In quantum theory, 
 unitarity and causality are related notions, and breaking of
unitarity leads usually to breaking of causality (see more
detailed discussion at the end of  Sect. 5). Causality
 in quantum gravity is broken also more directly via
production of {virtual} wormholes.}.
 In a quantum system with well--defined
Hilbert space endowed by a norm invariant under time evolution,
such a transformation of pure states into mixed states does not
happen, and nobody knows how to formulate a quantum theory where
the norm in  Hilbert space is not conserved.

To be more precise, there were attempts to formulate 
non--Schr\"odinger quantum theories. In the framework of the ADM
approach, one can naturally derive the so--called 
Wheeler -- de Witt equation \cite{WW}. It says
 \be
\label{WW}
\hat{H} \Psi \ =\ 0   
 \ee
(no term $i \dot{\Psi}$ on the right side). One obtains
zero on the right side, because the ADM Hamiltonian, the
generator of time translations, represents here
one of the gauge constraints: in gravity, the symmetry with
respect to coordinate translations is local, not global one.
There are comparatively 
``cosher'' quantum systems described by the 
wave equation of the Wheeler -- de Witt type. 
One of them is a quantum relativistic particle. 
The Klein-Gordon equation $(\hat{p}^2 - m^2)\Psi = 0$ has exactly
the form (\ref{WW}), and this is not accidental. The classical
action 
  \be
 \label{Spart}
S = \frac m2 \int \left(\frac {dx_\mu}{d\tau}\right)^2 d\tau
 \ee
is invariant with respect to reparametrizations $\tau \to
f(\tau)$ and reminds gravity in this respect. The
Klein
Gordon operator plays exactly the role of the ADM Hamiltonian.
However, this theory can also be formulated in a standard
Schr\"odinger form if choosing $x_0$ as time. The 
Schr\"odinger Hamiltonian is then $\hat{H}_{\rm Schrod}
= p_0 = \sqrt{\hat{\vec{p}}^2 + m^2}$. For the systems
with the wave function, which is changed not too rapidly 
(so that
the square root $\sqrt{-\partial_i^2 + m^2}\ \Psi $ is well 
defined), the equations $\hat{H}_{\rm ADM} \Psi \ =\ 0$
and $\hat{H}_{\rm Schrod} \Psi = i \dot{\Psi}$ are equivalent.   
  
For gravity, one can in principle also use this trick, but
 \begin{itemize}
\item
 Even for the simple system (\ref{Spart}), 
there is still no complete equivalence of the Schr\"odinger
equation and the Wheeler -- de Witt one; the restriction
for the wave functions  not to change too rapidly
should be imposed. Moreover, at least in the case when external
electromagnetic field is present, the Klein-Gordon equation (as well as the Dirac one) is known to be not internally self--consistent
because it does not take into account the creation of 
particle--antiparticle pairs, which always occurs in strong fields. 

\item In gravity (in contrast to the 
relativistic particle), we do not
have a unique natural recipe how to choose time. As a result,
the system (\ref{WW}) meets very serious, probably insurmountable
difficulties in interpretation \cite{Isham}.     
\end{itemize}

\section{String Story.}
Besides the difficulties discussed above, a standard 
quantum gravity also has another problem: it is a theory
with dimensional constant $m_P$ and, as such, is non-renormalizable.
This refers to the quantum version of the standard Einstein's 
gravity and also to its supersymmetric versions (though some 
divergences cancel out in supergravity, even ${\cal N} = 8$
supersymmetry is not powerfull enough to get rid of the infinite
number of counterterms).
   To cure this problem, string theory was  invented. The latter 
cures it by the simple reason: a finite size of a string serves
as an ultraviolet regulator and the ultraviolet divergences are 
effectively cut off.

\begin{figure}
   \begin{center}
        \epsfxsize=350pt
        \epsfysize=130pt
        \vspace{-5mm}
        \parbox{\epsfxsize}{\epsffile{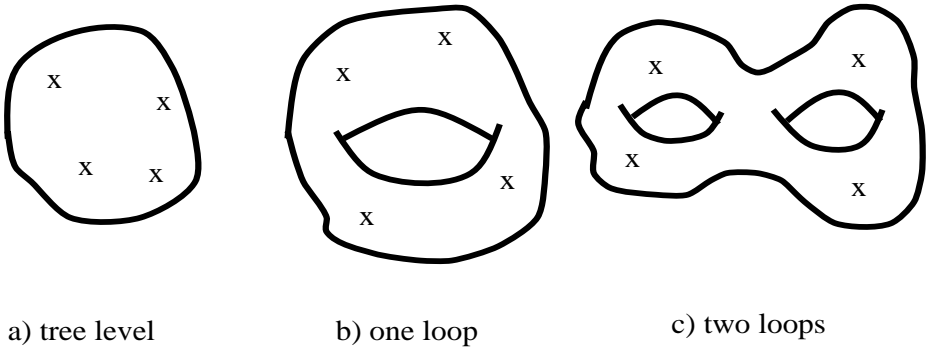}}
        \vspace{-5mm}
    \end{center}
\caption{String amplitudes. Crosses stand for sources.}
\label{petlistr}
\end{figure}

There are two points which I want to emphasize here.  
\begin{enumerate}
\item Even though perturbative string theory is, indeed, benign
in ultraviolet, it is in some sense {\it not constructed}
until now ! We understand it well at the tree level: we can
very well calculate tree string amplitudes described by the 
picture in Fig. 1a (2-sphere with sources) and also at the
1--loop level (torus with  sources). But already 
the calculation of the two--loop 
graph in Fig. 1c is a tremendously difficult
task. It involves integration over moduli space for 2-manifolds
of genus 2, and the latter has a very complicated structure. 
This moduli space (called Teichm\"uller space) 
involves certain singular points
corresponding to the cases when the width of one of
the handles in Fig. 1c shrinks to zero. 
The  integral for string amplitudes becomes divergent at these 
singular points, and though these divergences are not 
ultraviolet, but rather infrared in nature, they are
also nasty. Very recently, 
 the solution of the problem  for two--loop amplitudes
was announced \cite{2loop} (see also Ref.\cite{Danilov}), 
but we still do not
know how to treat divergences and calculate string
amplitudes in the general case.    

\item Even if consistent string perturbation theory for an 
arbitrary number of loops will ever be constructed, it will
solve the problem of renormalizability, but will hardly solve 
{\it real}
conceptual problems of quantum gravity discussed above: 
the absence of causality and unitarity. 

Let us discuss this point in some more details. String theory
has one nice feature compared to simple-minded quantum gravity:
if strings are embedded into {\it flat} multidimensional target
space
(usually called {\it bulk}), 
there {\it is} 
a natural definition of time. However, strings are nonlocal objects,
and, in the full theory treated nonperturbatively,
 this should bring about  noncausalities at 
Planck scale (though perturbative string amplitudes are
probably causal).
Noncausalities in the bulk are bound to lead to noncausalities in 
 effective 4--dimensional theory. 

 The question whether string theory is unitary has different answers
depending on whether we consider it in the bulk (then hopefully
it is) or from the viewpoint
of a 4--dimensional observer. In the latter case, it is definitely
not because the effective 4--dimensional theory is still Einstein's
(super)gravity and Hawking's paradox is still there. 

\end{enumerate}

My personal opinion (I will give more arguments in its favor
later) is that
string theory (at least the conventional string theory in
the framework of mid-eighties paradigma) has little chances to  
prove to be the fundamental theory of quantum gravity
and/or of Everything. Actually, nowadays most string
theorists also think that one should look {\it beyond} string
theory to find a really fundamental one ( {\it M--theory} ?).

My own suggestion, however, is that instead of
 looking {\it beyond} strings,  one can  try to
look in a different direction. 

\section{Conformal gravity}
 Going to strings instead of the fields is a rather bold and
radical step. The main conceptual problem is impossibility
to define Hilbert space and path integral in reasonably 
rigorous terms.
 
Of course, mathematicians maintain that the 
path integral is not
defined even in field theory, but for a physicist, there is no 
problem there. The Euclidean path integral is defined constructively
and has been calculated numerically by thousands of people 
since last 20 years. We {\it believe} that the Minkowski path
integral can also be calculated and the problem here is purely
technical. But for strings we have no idea how to do it. 
Whereas in field theory, we have an infinite number of dynamic
variables marked by spatial points $\vec{x}$, in
{\it string field theory}  dynamic variables are functionals
on loop space, i.e. the argument for the string field variable
is a particular 
 embedding
of the string in space $\{ \vec{x}(\sigma) \}$. In quantum
theory, the basic object would be a complex--valued 
``hyper--functional''
defined on the set of all such functionals. 
Many people tried to obtain some practical 
results in this direction, 
but to no avail. Two loops is the limit of
our understanding now.

Bearing this in mind, it is reasonable to explore less 
revolutionary approaches. String theory makes gravity 
renormalizable, but is it not possible to make it renormalizable
in a conservative field theory framework ?      

Yes, it is --- is the answer. Quantum version of Einstein gravity
is nonrenormalizable due to the presence of a dimensional constant.
It is easy to write a generally covariant Lagrangian where the
coupling is dimensionless and the theory is renormalizable.   
The Einstein-Hilbert action (\ref{SEinst}) is linear in $R$.
 Renormalizable
gravity is quadratic in $R$. There is a family of such theories 
with the actions
  \be
\label{SR2}
  S \ =\ \alpha \int  R_{\mu\nu} R^{\mu\nu}  \sqrt{-g} \ d^4 x \ +\ 
\beta \int R^2 \sqrt{-g} \ d^4 x \ .
  \ee
The structure $R_{\mu\nu\rho\sigma} R^{\mu\nu\rho\sigma} $
is reduced (at least, in the perturbation theory) to the two
structures in Eq. (\ref{SR2}) due to the Gauss--Bonnet identity
  \be
\label{GaBon}
R^2 -  4 R_{\mu\nu} R^{\mu\nu} + 
R_{\mu\nu\rho\sigma} R^{\mu\nu\rho\sigma}\ =\ {\rm total\ 
derivative\ .}
   \ee
It is known since long time that the theories of the class
(\ref{SR2}) are renormalizable. Moreover, they are asymptotically
free ! \cite{Fradkin}. We will concentrate on one particular 
theory in the family (\ref{SR2}) with the action
 \be
\label{Weylgrav}
  S \ =\ - \frac 1h  \int  
C_{\mu\nu\rho\sigma} C^{\mu\nu\rho\sigma}
\sqrt{-g} \ d^4 x \ ,
  \ee
where
 \be
\label{Weyltens}
C_{\mu\nu\rho\sigma} \ =\ R_{\mu\nu\rho\sigma} +
\frac 12 \left[ g_{\mu\sigma} R_{\nu\rho} + 
 g_{\nu\rho} R_{\mu\sigma} -  g_{\mu\rho} R_{\nu\sigma}
-  g_{\nu\sigma} R_{\mu\rho} \right] \nonumber \\
+ \ \frac R6 \left[  g_{\mu\rho}  g_{\nu\sigma} -
 g_{\mu\sigma} g_{\nu\rho} \right]
   \ee
is the Weyl tensor. A distinguishing feature of the theory
(\ref{Weylgrav}) is its invariance under {\it local} scale
transformations, 
  \be
  \label{scale}
g_{\mu\nu}(x) \ \to \ \lambda(x) g_{\mu\nu}(x)\ .
   \ee
 Bearing in mind the relation (\ref{GaBon}),
the action (\ref{Weylgrav}) is perturbatively equivalent 
to (\ref{SR2}) with $\beta = -\alpha/3 = 2/(3h)$. 

An immediate objection against the idea that the theory
 (\ref{Weylgrav}) describes the real
world could be that it does not have Newtonian limit.
Nonrelativistic potential corresponding to the action
(\ref{Weylgrav}) is not Coulomb-like, but 
grows  $\propto r$ (this follows from dimensional counting).
The objection to this objection is that {\it effective}
long--distance theory needs not to coincide with the fundamental
one. In fact, it can well coincide with Einstein's gravity !

As was mentioned, conformal gravity is an
asymptotically free theory. 
The explicit 1--loop calculation gives \cite{Fradkin}
 \be
\label{asfre}
\left. \frac 1h \right|_\mu\ =\ \frac 1{h_0} - \frac{199}{30}
\frac 1{16\pi^2} 
\ln \frac {\Lambda_{UV}} \mu \ ,
  \ee
where $\Lambda_{UV}$ is the ultraviolet cutoff. 
Asymptotic freedom  makes the physics of
conformal gravity rather similar to that of QCD. 
At large energies,
perturbation theory works, but at some scale $\mu \sim 
\Lambda_{\rm
Conf. \ Grav}$, where the effective
constant becomes large, nonperturbative effects come into play.
The scale $\Lambda_{CG}$ determines the mass of hadron--like
states. This is the standard dimensional transmutation. It is
natural to associate the scale  $\Lambda_{CG}$ with the Planck
scale.

In QCD, there are distinguished states, the pions, which remain
massless in the chiral limit. Thus, the effective theory for
massless QCD is the chiral theory describing pion interactions.
The form of the leading--order chiral effective Lagrangian
 \be
\label{chiral}
{\cal L}_{\rm chiral} \ =\ \frac {F_\pi^2}4  {\rm Tr}
\{\partial_\mu U \partial_\mu U^\dagger \} 
  \ee
is dictated by symmetry considerations.

The effective Lagrangian for conformal gravity is not invariant
under local scale transformations (\ref{scale}), but gereral
covariance should still be there. This dictates
  \be
\label{Seff}
  S^{\rm eff} \ =\ 
\Lambda \int   \sqrt{-g} \ d^4 x \ +\ 
\kappa \int R \sqrt{-g} \ d^4 x \ ,
  \ee
where $\Lambda$ is now cosmological term. 
{\it A priori}, $\Lambda \sim m_P^4$ and $\kappa
\propto m_P^2$. The estimate  $\Lambda \sim m_P^4$ is
about 130
orders of magnitude larger that the experimental value 
of cosmological constant. Thus, the theory (\ref{Weylgrav}) 
 is not viable as a realistic fundamental theory of gravity.
This refers actually to {\it any} nonsupersymmetric theory.
But if we start with supersymmetric conformal gravity without
cosmological term,
the induced cosmological constant vanish.
\footnote{In real world, supersymmetry is broken and it is
not clear, again, why cosmological constant is so small.
Nobody can answer now this troublesome question.}   
In other respects, the physics of conformal supergravity is 
similar to that of conformal gravity. In particular, conformal
supergravity is asymptotically free and involves dimensional
transmutation.

The second term in Eq. (\ref{Seff}) is the induced Einstein's 
gravity. The idea, by which
the Einstein--Hilbert action is not present in the tree action, but
is generated spontaneously due to  
 loops of usual matter fields was put forward long time ago by
Sakharov \cite{Sakharov}. It was mentioned in Ref.\cite{Zven} 
that this mechanism works also for conformal (super)gravity 
and the analogy with the dimensional transmutation mechanism
in QCD was emphasized.

At the scale $p_{\rm char} \sim \Lambda_{CG} \sim m_P$, 
nonperturbative effects come into play. In QCD, the nonperturbative
effects are not reduced to, but are well represented by instantons,
classical solutions to Euclidean field equations.
In gravity, there are also such solutions, they are
 Ricci--flat 4--dimensional
manifolds called gravitational 
instantons.  The simplest such solution is the 
Eguchi--Hanson solution
 \cite{EgHan} with the metric 
   \be
\label{EH}
ds^2 \ =\ \frac {dr^2}{1 - \frac {a^4}{r^4}} +
r^2\left[ \sigma_x^2 + \sigma_y^2 + \sigma_z^2 \left(1- 
\frac {a^4}{r^4} \right) \right]\ ,
  \ee
where
  \be
\label{sigmy}
\sigma_x &=& \frac 12 \left( \sin \psi d\theta - \sin \theta
\cos \psi d\phi \right) \nonumber \\
\sigma_y &=& -\frac 12 \left( \cos \psi d\theta + \sin \theta
\sin \psi d\phi \right) \nonumber \\
 \sigma_z &=& \frac 12 \left( d\psi  + \cos \theta d\phi \right) 
 \ee
are Cartan--Mauer forms. The metric (\ref{EH}) is locally 
asymptotically flat. It satisfies the condition $R_{\mu\nu} = 0$,
which are equations of motion for Einstein's gravity without
matter, but
Ricci flatness implies also that the equations of motion for
conformal gravity
 \be
\label{eqmotconf}
g_{\mu\nu} (3 C_{\alpha\beta\gamma\delta} 
  C^{\alpha\beta\gamma\delta} + 2 R^{;\alpha}_{\ ;\alpha} )
- 4 (R R_{\mu\nu} - R_{;\mu;\nu}) + \nonumber \\
12(2R_\mu^{\ \alpha} R_{\nu\alpha} - 
R_{\mu\nu\ \  ;\alpha}^{\ \  ;\alpha}
-    R_{\mu\alpha\beta\gamma} 
  R_\nu^{\ \alpha\beta\gamma}) \ =\ 0
  \ee
are satisfied. 

In constrast to Einstein's Euclidean action, which is not positive
definite and the corresponding path integral is ill--defined, the
Weyl action {\it is} positive definite. The Weyl action of Eguchi--Hanson
instanton is
 \be
\label{Sinst}
S^{\rm inst} \ =\ \frac {48\pi^2}h
 \ee
The contribution of the Eguchi-Hanson instanton to the path integral
is nonanalytic in $h$,  $\propto 
\exp\{- (48\pi^2)/h \}$, which is much similar to 
what happens in Yang--Mills theory. The EH instanton is analogous to the 
BPST
instanton also in other aspects: {\it (i)}
 The Riemann tensor for the EH instanton is
self--dual, as field strength tensor for BPST instanton is;
{\it (ii)} Like the BPST instanton, the EH instanton can be interpreted
as an Euclidean tunneling trajectory interpolating between two topologically
distinct vacua \cite{jainst}.
In the Yang--Mills case, different vacua are characterized by different
     Chern--Simons numbers. In the gravity case, there are
two classical vacua with  flat $R^3$
metric, but with different {\it orientation}. 
Following the EH instanton tunneling trajectory,
 flat $R^3$ space turns inside out and goes over to
 its mirror image. 

\newpage

\centerline{\Large Questions and answers.}

\vspace{.2cm}

Not everything is so rosy, however. Conformal gravity has also
 certain difficulties which we are in a position to discuss now.

First of all, when writing  Eq.(\ref{asfre}), 
we tacitly assumed (and this is true)  that the 
one--loop counterterm has the same functional form as the tree 
action. 
However, the classical conformal symmetry of the Weyl action
is broken by quantum effects. This means that we cannot guarantee 
that
higher--loop counterterms  are all proportional to 
(\ref{Weylgrav}). The
admixture of the structure $R^2$ cannot been ruled out. Thus, pure
Weyl gravity is not renormalizable. Of course, one
could consider the theory (\ref{SR2}) with two charges. 
Its physics
is roughly the same as  for the conformal gravity, but it is much 
less beautiful and hence much more suspicious.
The same concerns the ${\cal N} = 1$ supersymmetric version of 
Weyl theory.
It is also asymptotically free, conformal symmetry is anomalous, 
and nonconformal counterterms are bound to appear at the two loop 
level and higher. 

Aesthetically more appealing are the 
models where conformal symmetry of the 
classical action is sustained at quantum level. 
They are not only remormalizable, but simply
 finite: $\beta$ function
vanishes identically and counterterms of dimension 4 do not appear
whatsoever.
The most known example of such theory is
${\cal N} = 4$ supersymmetric Yang--Mills. Finite theories based
on conformal gravity are also known. The minimal variant of
  ${\cal N} = 4$ conformal supergravity happens not to
be finite,  but the coupling constant ceases to run
 if including an extra ${\cal N} = 4$
 SYM multiplet with the gauge group $U(1)^4$ or
$SU(2) \times U(1)$ \cite{Fradkin}.

If $\beta$ function vanishes, we do not have the mechanism of
dimensional transmutation at our disposal and the question arises
how the effective Einstein action involving a dimensional coupling
is generated. The answer is rather transparent: conformal symmetry
is not broken explicitly by quantum effects in  this case, but it 
can be broken {\it spontaneously}. The point is that
${\cal N} = 4$ finite theories involve scalar Higgs fields. For
certain nonzero values of the fields, classical potential vanishes. 
Supersymmetry dictates that the potential is not generated also
at quantum level: classical flat directions remain flat in quantum
theory. A set of all Higgs values where potential vanishes is called
{\it vacuum valley} or {\it vacuum moduli space}. This is a 
situation of neutral equilibrium: no particular point on the
vacuum moduli space is preferred, and we have a family of theories
characterized by particular Higgs expectation values.
This all is very well known for ${\cal N} = 4$ finite gauge 
theories, but it is also true for finite ${\cal N} = 4$ conformal
supergravities.

Higgs expectation values bring about dimensional
constants so that an effective low-energy theory is
 not conformal anymore. In the case  of finite gauge theories,
the effective theory is akin to the Standard Model, involving
spontaneous breaking of gauge symmetry by Higgs mechamism. 
The effective theory for the finite conformal supergravity involves
 Einstein's term and its superpartners. 

Let us discuss another difficulty  that conformal supergravity has.
The Lagrangian (\ref{SR2}) 
involves four derivatives of the metric. Field theories with
higher derivatives are usually considered sick because they
are intrinsically noncausal. The latter
 applies also to conformal gravity.
 To understand this, consider the theory involving on top of the
higher derivative terms also the Einstein term, 
${\cal L} \sim m_P^2 R + R^2$. The propagator of graviton
has then the form
\footnote{We are not worried  with numerical factors now.}
  \be
\label{propgrav}
D(k^2) \ \propto \ \frac 1{m_P^2 k^2 - k^4}  \ =\ 
\frac 1{m_P^2} \left( \frac 1{k^2} - \frac 1{k^2 - m_P^2} 
\right)\ . 
  \ee
In other words, on top of an ordinary massless graviton $G$, 
a massive
particle $G^*$ with {\it negative} residue at the pole appears. 
Production of particles with negative residues would violate 
unitarity. 

However, it is known  that unitarity is 
actually
not violated here  \cite{LeeWick,Tomb}. 
What {\it is} violated is causality.
The point is that, when loop corrections are taken into
account, the massive pole is shifted from the real axis,
the ``particle'' $G^*$ ceases to be an asymptotic state  and
cannot be produced in collision of usual massless gravitons. 
 Indeed,
nothing prevents the particle $G^*$ to go into a set of massless
gravitons, and this makes the polarization operator $\Pi(m_P^2)$
corresponding
to the propagator (\ref{propgrav}) complex. If $G^*$ were a 
``normal'' particle with positive metric, the resultant propagator
 $$\frac 1{k^2 - m_P^2 - 
\Pi(m_P^2)} $$
would involve a pole in the lower half-plane of $k_0^2$ (${\rm
Im} [\Pi(m_P^2)] < 0$ in this normalization).
When the residue is negative, the propagator
   $$\frac 1{-k^2 + m_P^2 - 
\Pi(m_P^2)} $$
has the pole in the upper half-plane of $k_0^2$ [$\Pi(m_P^2)$
is determined by the same graphs as for a usual particle 
and has the same value]. This property 
prevents making a usual Wick rotation
and is not consistent with causality. 
\footnote{In the  papers \cite{LeeWick}, 
higher derivative theories were studied mainly in association
with Pauli--Villars regularization procedure. The conclusion was
that the regularized Lagrangians lead to unitary amplitudes, but
that causality is broken at the regulator scale.}

Currently, it is not clear whether the causality breaking at Planck
scale persists
in the finite superconformal theories discussed above. In Ref. 
\cite{Fradkin} a careful optimism was expressed that may be it does
not. But even if it does, we do not see why it should be considered
as a major problem. At nonperturbative level, 
microcausality is  broken in {\it any} 
gravity theories, with string theory not presenting an exception.
In conformal supergravity models it is probably also broken 
 perturbatively.

So what ?

\section{Supergravity as a theory of 3--brane: \\ 
Ogievetsky--Sokatchev
approach.}

In the previous section, we argued that conformal supergravity
(probably, a finite, anomaly--free version thereof) can be
considered as a viable candidate for the fundamental gravity
theory. It solves the problem of nonrenormalizability of
standard gravity even better than string theory does 
(we say {\it better},
 because  perturbative calculations to any order in
coupling constant present no essential 
technical difficulties there)
and the difficulties it has are intrinsic for {\it any}
gravity theory.

String theory has one attractive feature, however. It is 
formulated not in  curved 4--dimensional space, but in
the flat multidimensional bulk. This gives a principle
solutions to the problem of time, and brings forward hopes
to construct self--consistent quantum theory.

We want to notice here 
that similar hopes can actually be associated
with {\it standard supergravity} if describing the latter
in the superfield formalism due to Ogievetsky and Sokatchev
\cite{OgSok}. 

Ogievetsky--Sokatchev approach to supergravity has a lot
of advantages compared to the standard Wess--Zumino approach.
Unfortunately, the former is not so widely known,
and we are in a position to explain briefly its basic features.
In Wess--Zumino approach, the basic superfield 
is $E^A_M$, a supersymmetric generalization of vierbein. 
This superfield has a lot of unphysical components; to get 
rid of them, one has to impose constraints of a rather
complicated form.

The Ogievetsky--Sokatchev approach is based on a beautiful
geometric construction. Consider a curved $(4+4)$ -- dimensional
supermanifold (it has 4 bosonic coordinates $x^m$
and 4 real or 2 complex fermionic coordinates $\theta_\alpha$)
embedded into  {\it flat} $(8+4)$ -- dimensional superspace 
involving 4 {\it complex} (which is equivalent to 
8 real) bosonic coordinates $z^m$ 
and 2 complex fermionic coordinates.  
Such an embedding is characterized by the superfield 
${\cal H}^m(x^n; \bar\theta^{\dot{\alpha}}, \theta_\alpha)$, where
${\cal H}^m$ coincides with the imaginary parts of flat coordinates
$z^m$ and $x^m$ --- with their real parts. The Lagrangian
of the standard Einstein supergravity is none other than
 the supervolume of the associated hypersurface:
  \be
\label{sdet}
S_{\rm sugra} \ =\ m_P^2 \int {\rm Ber} \    \|
 E_M^A  \|  
\ d^4 x \ d^4 \theta \ ,
  \ee
where 
$E_M^A$ is the {\it induced} super--vielbein on the 
hypersurface and ``Ber'' stands for the Berezinian 
(or superdeterminant).
 Now, $E_M^A$  and ${\rm Ber} \|E \|$
can be expressed in terms of
${\cal H}^m(x^n; \bar\theta^{\dot{\alpha}}, \theta_\alpha)$
(in a not so simple, but explicit way). One can check that they
obey the coinstraints that are imposed on $E_M^A$ in the 
Wess--Zumino approach. On the other hand,    
 no constraints on the {\it axial superfield}  
(we are using the Ogievetsky--Sokachev terminology) ${\cal H}^m$ 
need be imposed.   

The Lagrangian (\ref{sdet}) 
 is invariant with respect to  general 
reparametrizations of all bosonic and fermionic coordinates
on the hypersurface. 
This group is too  large, however, which is not convenient. 
In addition,  a generic such reparametrisation destroys the simple
form 
  \be
  \label{formH}
{\rm Im} (z^m) \ =\ {\cal H}^m \left({\rm Re}(z^n), \bar\theta,
\theta \right)
  \ee 
chosen by us to describe the hypersurface. 

The form (\ref{formH}) 
is preserved by a subgroup of the general reparametrisation group.
To describe it, introduce
left and right coordinates $x^m_{L,R} = x^m \pm
i{\cal H}^m$ and require them to reduce to the familiar
 \be
\label{xLR}
x^m_{L,R} \ =\ x^m \pm i \bar \theta \sigma^m \theta \ .
  \ee 
in the limit when the embedded hyper--surface represents
a hyper--plane.
Then the  transformations
 \be
\label{reparam}
x^m_L &\to& f^m(x^n_L, \theta_\beta)
\nonumber \\
\theta_\alpha &\to& \chi_\alpha 
(x^n_L, \theta_\beta)
  \ee
obviously preserve the form (\ref{formH}). To provide for the 
invariance of the action (\ref{sdet}) or, which is the same,
to provide for that the transformations (\ref{reparam})
represented a reparametrization of the coordinates on the 
hypersurface, it is sufficient to require that the 
super-Jacobian of the transformation (\ref{reparam}) be
equal to 1,
 \be
 \label{supJac}
{\rm Ber}  \left \| \frac {\partial(x_L', \theta')}
{\partial(x_L, \theta)}\right\|  =
\det \left \| 
\frac {\partial x^{'m}_L}{\partial x^n_L}
- \frac {\partial x^{'m}_L}{\partial \theta_\alpha} 
\frac {\partial \theta_\alpha}{\partial \theta'_\beta}
\frac {\partial \theta'_\beta}{\partial x^n_L}   \right\|
\det^{\ \ \ \ \ \ -1}  \left \| 
\frac {\partial \theta'_\alpha} {\partial \theta_\beta}
 \right\|   = 1\ . 
    \ee
The gauge symmetry (\ref{reparam}) allows one to greatly reduce
the number of components of ${\cal H}^m$. 
There are all together
64 components. The transformations (\ref{reparam}) involve 
48 parameters, but the condition (\ref{supJac}) fixes 8 of them
leaving 40 free parameters. As a result, we obtain 24 
(12 bosonic and 12 fermionic)
gauge--invariant degrees of freedom. They exactly correspond
to component language counting  \cite{Niven}. The 
Lagrangian involves 38 components
(16 for the vierbein $e^m_a$, 16 for the gravitino 
$\psi_\alpha^m$, and 6 for the auxiliary fields $S, P, A^m$.
There are 4(general coordinate) plus 6(local Lorentz) plus
4(supersymmetry) = 14 gauge parameters. Now, $38-14 = 64 - (48-8)
 = 24$.   

It is convenient to choose the {\it normal gauge} (analogous to the 
Wess--Zumino gauge used in the analysis of supersymmetric
gauge theories), in which case
 \be
\label{WZgauge}
 {\cal H}^m \ \sim \ 
e^m_a \bar\theta \sigma^a \theta + {\rm other\ terms}\ .
 \ee
One can then be directly convinced 
(though the calculation {\it is} tedious)
that the bosonic part of the action (\ref{sdet}) 
coincides (up to a total
derivative !) with $R$. The other terms in the component
Lagrangian are  restored by supersymmetry.

Now, ${\cal N} =1$ conformal supergravity can also be described
in these terms: its Lagrangian can be expressed via the 
unconstrained
axial superfield ${\cal H}^m$. This Lagrangian (see the papers
\cite{OgSok} for explicit formulae) is invariant with respect
to the general transformations (\ref{reparam}) (not restricted
by the requirement of unit super--Jacobian ). 

Note in passing
that also the variant of supergravity with cosmological term
is  nicely expressed in the Ogievetsky--Sokatchev formalism.
It turns out that the corresponding action represents a 
{\it total\ derivative} and the problem is reduced to the
choice of boundary conditions. Thus, the question why  the 
cosmological
term vanishes acquires the same status as the question
why the $\theta$ term in QCD vanishes. 
No comprehensive answer to any of
these questions is known, but we are sure at least that, if we
start with a supersymmetric theory with 
vanishing cosmological term, the latter
 is not generated by quantum 
effects, by the same token as the $\theta$ term in QCD 
is not generated.  

Our main point is that, 
once flat space appeared in the formulation of the theory, 
a natural definition of  time exists, which should allow one 
to present the equations of motion in the Causchy form.
The theory becomes much similar to string theory, only it is
in a sense much 
more complicated: the latter deals with embeddings of 2--surfaces
into flat Minkowski
 bulk, while the former depends on embeddings of
4-surfaces (3--branes in modern terminology) there.

On the other hand, supergravity is still much {\it simpler} than the
full string field 
theory. Indeed, in spite of the fact that the action
(\ref{sdet}) describes multidimensional geometry, it is
{\it four--dimensional} in nature. The basic dynamic variables 
in such theory are embeddings themselves rather
than frightening functionals in the loop space, which we would
eventually have to learn to deal with if sticking to the
conventional string theory paradigma.

\section{Discussion.} 
 Before going further, let us  reiterate
briefly the main points of our reasoning so far
(you may call it {\sl party
line}, bearing in mind that the corresponding party is not
numerous and in opposition).

\begin{enumerate}

\item We do not know how to construct a consistent gravity
theory strictly in a four--dimensional framework. The main
problem here is the problem of time, which has not been fully
solved even in classical general relativity
 and becomes a real mayhem
when one attempts to quantize it.

\item Quantum version of Einstein's gravity has another problem:
nonrenormalizability. It persists in supersymmetric 
generalizations. 

\item The latter problem is cured in string theory, 
but a {\it simpler} and in many respects { nicer} medicin
is provided by conformal gravity. The {\it effective}
low--energy theory for conformal gravity is 
Einstein's theory (modulo the problem of cosmological term,
which is more tractable for supersymmetric versions of
the theory, but is far from being fully resolved).
In a nonsupersymmetric
or ${\cal N} =1,2,3$ supersymmetric versions of the theory
involving conformal anomaly, 
Einstein's constant is generated due to dimensional 
transmutation mechanism. We like better ${\cal N} = 4$ finite
superconformal theories, where  Einstein's constant is generated
due to spontaneous breaking of conformal symmetry when a particular
point on flat Higgs moduli space is picked up.

\item With all probability, causality is broken in these
theories at  perturbative level (though this was not explicitly
demostrated) due to the presence of higher derivatives in the
Lagrangian and complexification of negative metric poles by
Lee and Wick mechanism. But {\it any} gravity theory is acausal
in four dimensions. 

\item ${\cal N}= 1$ supergravity and conformal supergravity have
a nice interpretation due to Ogievetsky and Sokatchev, where
the classical field configuration can be thought of as an embedding
of a 3--brane into 8--dimensional flat bulk space. This gives
one a natural definition of time, and one can hope to construct
a unitary  quantum theory with well--defined Hilbert 
space {\it in the bulk}. The reasons
are the same that the reasons why we believe that  string theory
(we mean string theory in the  second
quantization  framefork, when it is a form of
2--dimensional field theory) is unitary  in the bulk. 
 
\end{enumerate}

As the reader has probably already guessed, we {\it believe}
\footnote{As we live now in civilized times and the risk of 
being severely
punished (beaten by stones, etc) for a false prophecy is 
comparatively low, I am allowing myself to make one.}
that the future fundamental theory of gravity (and probably
of Everything) is a variant of finite superconformal gravity
theory. We also believe that this theory can be represented
as a theory of 3--brane embedded into a higher--dimensional
flat space.

There are still several points which are not clear now.
The last one is especially worrysome.
\begin{enumerate}

\item We {\it believe} that Ogievetsky--Sokatchev supergravity
is unitary and causal in the bulk, but do not know how to prove
it. This is going to be much more complicated problem than
proving unitarity for string theory (such a {\it proof} is
also absent now).

\item Nice geometric interpretation discussed above has been found
so far only for ${\cal N} =1$ theories. 
Little is known in this respect
about ${\cal N} = 4$ theory. An educated guess is that the bulk
is this case is 10--dimensional rather than 8--dimensional.
One can notice in this respect that, in the problem of embedding
of a 4--dimensional manifold into $R^m$ , the dimensions 8
and 10 are distingushed. Namely, {\it (i)} one can always embed an
$n$--dimensional manifold into $R^{2n}$ without self--crossings
and {\it (ii)}
one can always embed  an
$n$--dimensional manifold into $R^{2n+2}$ without knots (so that
 all embeddings of a given manifold are topologically
equivalent) \cite{embed}.

\item The finite superconformal gravity theories discussed above
do not have realistic matter content. They are based on the
gauge group $SU(2) \times U(1)$ or $U(1)^4$, whereas we need
the group $SU(3) \times SU(2) \times U(1)$ or larger, 
three fermion generations, etc. 
It is not clear, however, that realistic
superconformal gravity theories will never be found. 

\item There is also a major philosophical problem. The physics
of 20-th century is based on  positivistic philosophy.
We want to formulate theory in terms of physical observables
and dismiss as meaningless all attempts to talk about ``real''
electron trajectories, etc. A real physical observable is by 
definition something which can be measured by a real physical 
observer,
who is four--dimensional, like we are. But if we treat the theory 
in a multidimensional bulk, the wave function of the Universe and 
(in the proposed approach) the $D3$--brane transition amplitudes
can be measured only by a ``divine'' observer living in the bulk.
This smells mystics, but I do not know how to get rid of it here. 

\end{enumerate} 

\begin{figure}
   \begin{center}
        \epsfxsize=300pt
        \epsfysize=450pt
        \vspace{-5mm}
        \parbox{\epsfxsize}{\epsffile{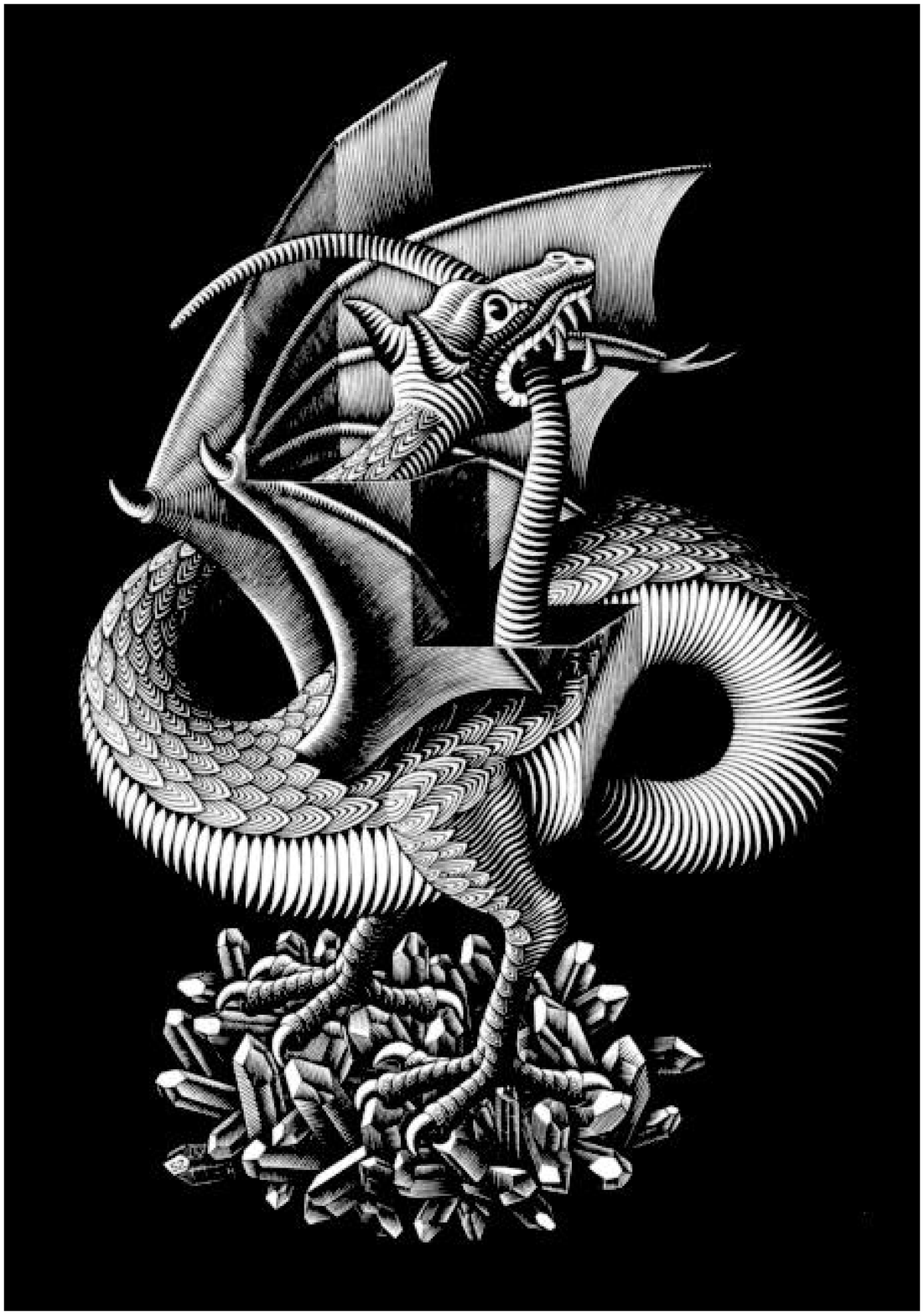}}
        \vspace{-5mm}
    \end{center}
\caption{Dragon.}
\label{dragon}
\end{figure}

For any oppositioner, the negative program is usually much stronger
than the positive one. That is why I want to finish with some
comments on what, I think, the fundamental theory of Everything
is {\it not}. 

I am personally rather skeptical towards the assertions
 that higher
dimensions are {\it really} there. The matter is that even ordinary
field theory is ill--defined if the dimension of space--time is
5 or more: the path integral simply does not have continuum limit 
there (at least, for $D \geq 5$, we are not aware of any example
where such a limit existed). 
I cannot imagine that the string field theory path integral
is defined any better. Thus, I do not believe in the ideas 
(rather
popular now) of large extra dimensions, the brane new world, etc.

It may be beneficial and even necessary to think of our physical
space as being embedded into a multidimensional flat bulk, but
the physical space itself should be four--dimensional. In other
words, my attitude towards higher dimensions
 is close to the standpoint of Catholic 
Church with respect to  heliocentric 
ideas of Kopernicus and Galileo. 
No problems as far as they were proposed as a convenient 
mathematical tool to facilitate calculation of physical 
observables like planet positions, etc
(for people of 16-th century, the physical observer
must, of course, dwell on Earth), but the suggestion
that Earth {\it really} rotates around Sun was unacceptable.
\footnote{My reasons are not religious, however, but simply
a desire to be able to define the path integral.}

Close to the end, but not in the very end, 
I want to present, on top of a historico-philosophical analogy,
an artistic one  and  simultaneously justify
the queer title of this paper.
``{\sl Dragon}'' is a gravure by Escher. It is reproduced
 in Fig. \ref{dragon}.
As was emphasized in Ref.\cite{Hof}, 
this dragon seems to be very much three--dimensional,
it kind of tries to escape the sheet of paper where it is drawn.
But the only ``physical dragon'' that is at our disposal is the
gravure itself, which is two--dimensional. It tries to make us
believe that his real dimensionality is more than two, but it is
a {\it false} claim. Likewise, gravity 
may be conveniently formulated
in higher--dimensional terms, but 
our physical world has only 4 dimensions.

The last paragraph of the paper is reserved to a physical 
argument. The idea that an essentially
 four--dimensional theory can be conveniently described with
fictitious 
higher--dimensional scaffolds is not new. This is exactly the
content of Maldacena's conjecture on AdS/CFT correspondence:
the correlators of four--dimensional SYM theory coincide with 
certain
correlators in 10--dimensional supergravity defined on the 
boundary of some particular background \cite{Malda}. 
Many other quantities in 10--dimensional theory can be defined
and considered, but they are declared to be meaningless as
far as SYM theory is concerned.

\section*{Acknowledgements}
I am indebted to Sergei Blinnikov, Ian Kogan, and
Arkady Tseytlin  for illuminating discussions.

\end{document}